\documentclass[aps, prl, twocolumn,tightenlines]{revtex4}
\usepackage{epsfig}
\usepackage{amsmath}
\usepackage{dcolumn}
\usepackage{bm}

\begin{document}

\title{Nonlinear interferometric vibrational imaging}

\author{Daniel L. Marks}
\affiliation{Beckman Institute for Advanced Science and
Technology, University of Illinois at Urbana-Champaign}
\author{Stephen A. Boppart}
\email{boppart@uiuc.edu} \affiliation{Department of Electrical and
Computer Engineering, Bioengineering Program} \affiliation{College
of Medicine} \affiliation{Beckman Institute for Advanced Science
and Technology, University of Illinois at Urbana-Champaign, 405
North Mathews Avenue, Urbana, IL 61801}

\date{\today}

\begin{abstract}
Coherent Anti-Stokes Raman Scattering (CARS) processes are
``coherent,'' but the phase of the anti-Stokes radiation is usually
lost by most incoherent spectroscopic CARS measurements.  We propose a
novel Raman microscopy imaging method called Nonlinear Interferometric
Vibrational Imaging, which measures Raman spectra by obtaining the
temporal anti-Stokes signal through nonlinear interferometry.  With a
more complete knowledge of the anti-Stokes signal, we show through
simulations that a high-resolution Raman spectrum can be obtained of a
molecule in a single pulse using broadband radiation.  This could be
useful for identifying the three-dimensional spatial distribution of
molecular species in tissue.
\end{abstract}

\maketitle

Functional imaging techniques have been developed to
provide insight into biological processes.  Optical functional
imaging is frequently limited because dyes or markers must
be introduced that alter or damage biological tissues.  Because it is
preferable to use endogenous properties of tissues to identify
molecular contents, methods such as infrared and Raman spectroscopy
are used.  In particular, Coherent Anti-Stokes Raman Scattering (CARS)
processes have been successfully integrated with confocal scanning
microscopes~\cite{duncan1982, potma2002,
cheng2001-2} to achieve three-dimensional molecular
images of vibrational resonances.  However, existing instruments
measure only the total power of the received anti-Stokes radiation.
We propose a novel method, Nonlinear Interferometric Vibrational
Imaging (NIVI), which utilizes nonlinear interferometry to measure the
amplitude and phase of the anti-Stokes light.  An experimental
demonstration of the principle of this technique has been
demonstrated.~\cite{bredfeldt2003}.  This additional phase
information facilitates the inference of the amplitude and phase of
the nonlinear susceptibility of the molecule.  By utilizing NIVI with
properly designed illuminating radiation, a large region of the
amplitude and phase of the Raman spectrum can be sampled in a single
brief pulse.

Coherent Anti-Stokes Raman Scattering processes have only recently
been used to probe biological specimens.  The appeal of CARS is that
it can probe the density of molecules with a particular Raman
resonance frequency while exposing the specimen to relatively low
levels of illumination.  A typical CARS process illuminates the
specimen with a pump pulse of frequency $ \omega_1 $, and a Stokes
pulse of frequency $ \omega_2 $, which are separated by the
vibrational frequency of interest $ \Omega=\omega_1-\omega_2$.  If a
molecule with a Raman resonance at frequency $ \Omega $ is present, an
anti-Stokes pulse at frequency $ 2\omega_1 - \omega_2 $ is produced.
In CARS microscopy, tightly focused pump and Stokes beams are scanned
through the specimen, and the anti-Stokes photon count is measured at
each point.  This photon count is proportional to the square of the
molecular bond density and the magnitude of the Raman susceptibility.
Nonlinear interferometry has been used to characterize the magnitude
of the stimulated Raman scattering nonlinearity in
liquids~\cite{owyoung1977}, and also of two CARS
signals~\cite{hahn1995} in gases.

The model of CARS with broadband pulses used here is similar to that
described by~\cite{oron2002,oron2002-2}.  We do not assume that the
illuminating radiation is narrowband.  However, we stipulate that the
molecule does not resonantly interact directly with any of the
frequencies inside the illumination bandwidth or the generated
anti-Stokes bandwidth of the optical signal.  The CARS process is
composed of two stimulated Raman scattering processes involving four
photons that begins and ends with the molecule in the ground state.
To describe CARS, we denote the electric field incident on the
molecule as $ \tilde{E_i}(\omega) $.  The first process is modeled by
Eq.~\ref{eq:cars1}, and excites the nonlinear dipole polarization of
the resonant transition.
\begin{eqnarray}
\label{eq:cars1}P^{(3)}(\Omega) =  \chi^{(3)}(\Omega)\int\limits_0^\infty{\tilde{E_i}(\omega+\Omega)\tilde{E_i}(\omega)^*\ d\omega} \mbox{ (step 1)}\\
\label{eq:cars2a}\tilde{E_o}(\omega) = \int\limits_{0}^\omega{\tilde{E_i}(\omega-\Omega)P^{(3)}(\Omega)\ d\Omega} \mbox{ (step 2)}
\end{eqnarray}
Each pair of frequencies that are separated by a resonance of the
molecule at frequency $ \Omega $ produces a nonlinear polarization in
the molecule.  Another way to look at Eq.~\ref{eq:cars1} is that, in
the time domain, the molecule has a nonlinear polarization that is
driven not by the electric field but by the instantaneous intensity
envelope of the signal.  In this formulation, we are neglecting any
changes in $ \chi^{(3)} $ that are dependent on the
``carrier'' envelope frequency on which the beats are imposed.
Therefore any pulse train with intensity beats at the resonance
frequency will stimulate the nonlinear polarization.  Examples of this
are pulses that are modulated periodically in the spectral domain with
period $ \Omega $
~\cite{oron2002,oron2002-2,dudovich2002}, and
interfering two relatively delayed chirped pulses to achieve a beat
frequency proprortional to time delay~\cite{gershgoren2003}.
The second step creates anti-Stokes radiation by mixing the incoming
radiation field with the polarization in the time domain, and is
modeled by Eq.~\ref{eq:cars2a}.  The Eqs.~\ref{eq:cars1}
and~\ref{eq:cars2a} allow one to calculate the emitted CARS radiation
$ \tilde{E_o}(\omega) $ for a given $ \tilde{E_i}(\omega) $ and $
\chi^{(3)}(\Omega) $.  While these relations do not constitute a
linear relationship between $
\tilde{E_i}(\omega) $ and $ \tilde{E_o}(\omega) $, there is a linear
dependence of $ \tilde{E_o}(\omega) $ on $ \chi^{(3)}(\Omega) $ given
an input field $ \tilde{E_i}(\omega) $.  This suggests that with
knowledge of the complex $ \tilde{E_o}(\omega) $, one can do linear
estimation to find $ \chi^{(3)}(\Omega) $.  The advantage of NIVI over
incoherent detection is that nonlinear interferometry enables the
recovery of the complex $ \tilde{E_o}(\omega) $.  With a
properly designed input pulse $ \tilde{E_i}(\omega) $, the nonlinear
susceptibility can be found in a particular frequency range.

NIVI takes advantage of the coherent nature of the CARS process to
allow the phase-sensitive measurement of the anti-Stokes radiation.
Conventional linear interferometry involves splitting a source light
beam into two parts, each of which scatters linearly in the field, and
which are then recombined and detected.  NIVI
differs in that a CARS process occurs to one of the split beams.

\begin{figure}
\epsfig{figure=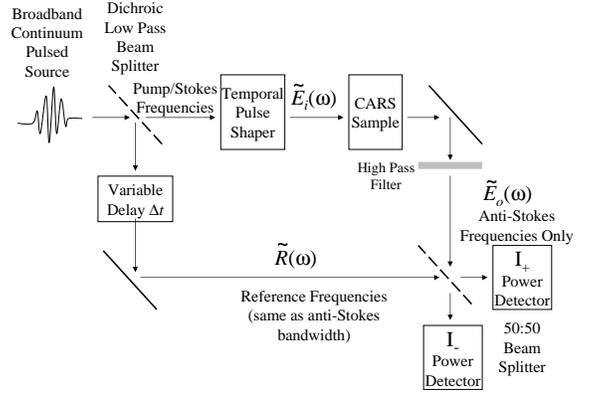,width=90mm,bbllx=0,bblly=200,bburx=761,bbury=700}
\caption{Schematic of NIVI with a broadband continuum source,
using the continuum for the reference pulse.  The detection scheme
is simplified and in practice a single-shot detection scheme such
as~\cite{purchase1993} could be used.} \label{fig:fig1}
\end{figure}

NIVI can be implemented with the setup detailed in
Fig.~\ref{fig:fig1}.  We start with a broadband, phase-locked
source of light pulses, such as those from a mode-locked laser.
Sources that can produce such light are ultrabroadband mode-locked
Ti-sapphire oscillators~\cite{drexler1999}, and supercontinuum
sources~\cite{wadsworth2002, marks2002}.  Because
the source is phase-locked, there is a deterministic relationship
between phases at various frequencies.  This deterministic
relationship will be preserved by coherent processes such as CARS.  To
utilize this determinacy, the source bandwidth is split into higher
and lower frequency bands, as shown in Fig.~\ref{fig:fig1} with a
dichroic beamsplitter.  The higher frequency band will be a reference
signal $\tilde{R}(\omega)$, and will correspond to the bandwidth of
the anti-Stokes frequencies produced by a sample.  The lower frequency
band is temporally shaped to stimulate CARS in the sample, a signal we
denote by the frequency spectrum $ \tilde{E_i}(\omega) $.  Some of the
illumination signal will be converted to anti-Stokes radiation by the
sample.  Because CARS processes are usually rather weak, we will
assume that any new anti-Stokes radiation created in the same
bandwidth as the illumination will be inseparable from the
illumination. Therefore we will discard all anti-Stokes light inside
the illumination bandwidth with a high-pass frequency filter.  The
remaining anti-Stokes light that passes through the filter, which we
denote by the spectrum $
\tilde{E_o}(\omega) $, corresponds to frequencies in the reference
signal.  We combine the reference signal $ \tilde{R}(\omega) $ and the
anti-Stokes spectrum $
\tilde{E_o}(\omega) $ with a 50:50 beam splitter, and utilize balanced
detection to measure the intereference component on two
photodetectors.  There is a delay of time $ \Delta t $ placed in the
reference path to facilitate measuring the temporal cross-correlation
between the reference and anti-Stokes signals.  The difference in
the two intensities $
\Delta I(\Delta t) $ as a function of delay between the signal and
reference will be:
\begin{equation}
\Delta I(\Delta t)= I_+ - I_- = \int\limits_{0}^\infty{ 4\,\mbox{Re}\left\{\tilde{E}_o(\omega)^*\tilde{R}(\omega)\exp(i \omega \Delta t)\right\}\ d\omega}
\label{eq:intf2}
\end{equation}
If we call $ \tilde{I}(\omega) $ the Fourier transform of $
\Delta I(\Delta t) $ with respect to $ \Delta t $, we find that
$ \Delta \tilde{I}(\omega) = 4\, \tilde{E}_o(\omega)^*\tilde{R}(\omega) $.
Thus the measured data retains its linear relationship with respect
to the anti-Stokes spectrum  $ \tilde{E}_o(\omega) $ and therefore
the nonlinear susceptibility $ \chi^{(3)}(\Omega) $.

Besides the ability to find the complex-valued $ \chi^{(3)}(\Omega) $,
interferometry eliminates the need for photon-counting detectors.
Another advantage is that interference will only occur when the
anti-Stokes light and the reference light arrive at the beam splitter
at the same time.  Because of this, temporal gating can be used to
produce three-dimensional vibrational images in a manner analogous to
Optical Coherence Tomography~\cite{huang1991,boppart1998,bouma2001}.
Coherent detection is also far less sensitive to stray light than
photon counting.  Because of this, NIVI may be more adaptable to
various scanning configurations and environments outside the
laboratory.

To show that NIVI can measure intervals of the Raman spectrum in a
single pulse, a pulse must be designed that can stimulate molecules in
a broad Raman spectrum.  The approach we take creates beats that
instead of being of a constant frequency~\cite{gershgoren2003}, will
be themselves chirped.  This can be accomplished by combining two
chirped pulses with a relative delay, but with different chirp rates.
If we have a transform-limited pulsed source of center frequency $
\omega_0 $ and bandwidth $ \Delta
\omega $, and we wish to sweep the beat frequency from $ \Omega_L $ to
$ \Omega_H $ in time $ T $, we can design a pulse $
\tilde{E_i}(\omega) $ such that:
\begin{widetext}
\begin{equation}
\begin{array}{l}
\tilde{E_i}(\omega) = E_0 \cos\left(\frac{\pi(\omega-\omega_0)}{\Delta \omega}\right) \left[\left(\frac{1+\kappa}{2}\right)\exp\left(\frac{-i (\omega-\omega_0) \tau}{2} - \frac{i (\omega - \omega_0)^2}{2(\alpha+\beta)}\right) + \right. \\
\ \ \ \ \ \ \ \ \ \ \ \left(\frac{1-\kappa}{2}\right)\left.\exp\left(\frac{ i (\omega-\omega_0) \tau}{2} - \frac{ i (\omega - \omega_0)^2}{2(\alpha-\beta)}\right)\right]\mbox{ for }\omega_0-\frac{\Delta \omega}{2}<\omega<\omega_0 + \frac{\Delta \omega}{2} \\
\tilde{E_i}(\omega) = 0 \mbox{ otherwise } \\
\ \ \ \ \ \ \ \ \ \ \mbox{where } \alpha = \frac{2 \Delta \omega - \Omega_H - \Omega_L}{2T}, \beta = \frac{\Omega_H - \Omega_L}{2T}, \mbox{ and } \tau = \frac{T}{2}\left(\frac{\Omega_H}{\Delta \omega-\Omega_H}+\frac{\Omega_L}{\Delta \omega-\Omega_L}\right)
\end{array}
\label{eq:nivi2}\end{equation}
\end{widetext}
The variable $ \alpha $ is the common chirp to both pulses, $ \beta $
is the difference chirp, $ \tau $ is the time delay between the two
pulses, and $ \kappa $ is the difference in field magnitude between
the two pulses.  The pulse bandwidth has been apodized with a cosine
window because in practice it seems to help the stability of the
inversion.  Note that the bandwidth of the source $ \Delta \omega $
must exceed $ \Omega_H $ so that beats can be formed at all Raman
frequencies.  When creating the pulse, the chirp time $ T $ will
control the resolution with which one will be able to resolve
frequencies in the Raman spectrum.  The largest practical $ T $ is
determined by the dephasing time of the resonances, which in most
liquids is on the order of picoseconds.

To demonstrate the feasibility of NIVI, we simulate the illumination
of a target molecule with the broadband pulse of Eq.~\ref{eq:nivi2}
and use the returned signal to estimate the complex susceptibility $
\chi^{(3)}(\Omega) $.  We will show two simulations: one that is able
to probe a wide bandwidth of Raman resonances in a single pulse, and
the other which is able to distinguish between two nearby resonances.
We take as our hypothetical laser source a mode-locked Ti-sapphire
laser that can produce a pulse with a uniform bandwidth from 700--1000
nm, and the setup of Fig.~\ref{fig:fig1}.  The bandwidth from
800--1000 nm will be reserved for stimulating CARS, with the remainder
used as the reference signal.  For the first simulation, the CARS
excitation bandwidth will be shaped such that $ \Omega_L = 700 \mbox{
cm}^{-1} $, $ \Omega_H = 1300 \mbox{ cm}^{-1} $, and $ T = 5 \mbox{
ps} $.  To show that the system can reconstruct several simultaneous
resonances over the entire bandwidth, we create a hypothetical $
\chi^{(3)}(\Omega) $ with several Lorentzian resonances centered at
$800 \mbox{ cm}^{-1}$, $900 \mbox{ cm}^{-1}$, $1000 \mbox{ cm}^{-1}$,
and $1100 \mbox{ cm}^{-1}$.  These frequencies are in the Raman
``fingerprint'' region and would be useful for practical molecular
identification.

The simulation was implemented by sampling the spectra of $
\chi^{(3)}(\Omega) $, $ \tilde{E_i}(\omega) $, and $
\tilde{E_o}(\omega) $ with 20,000 points spaced at equal intervals
from $ 0 \mbox{ cm}^{-1} $ to $ 20000\mbox{ cm}^{-1} $ in $ 1.0 \mbox{
cm}^{-1} $ steps.  The cross-correlations of Eqs.~\ref{eq:cars1}
and~\ref{eq:cars2a} were computed using the Fast Fourier Transform.
These two equations form a ``forward'' CARS linear operator computing
$ \chi^{(3)}(\Omega) $ from $ \tilde{E_o}(\omega) $ which we call $ {\bf
A}(\omega,\Omega) $.  To find the inverse of this operator, we used
the Tikhonov-regularized least-squares inversion operator, which is
formally denoted by $ {\bf A}^* = ({\bf A}^\dag {\bf A} + \epsilon
{\bf I})^{-1} {\bf A}^\dag $.  The Tikhonov regularization was
included to improve the stability of the inverse and to account for
potential noise sources such as thermal and photon noise in a
practical experiment.  The constant $ \epsilon $ is chosen to account
for the magnitude of additive white Gaussian noise in a realistic
experiment.  In practice, $ {\bf A}^* $ was computed using the
preconditioned conjugate gradient~\cite{golub1996} method to avoid the
very computationally expensive direct matrix inversion.  While we do not model a
real noise source here, Tikhonov regularization adjusts the inverse
operator such that features of the estimated spectrum
$ \chi^{(3)}(\Omega) $ that would be unstable due to
insufficient information for reconstruction would tend towards zero.

\begin{figure}
\epsfig{figure=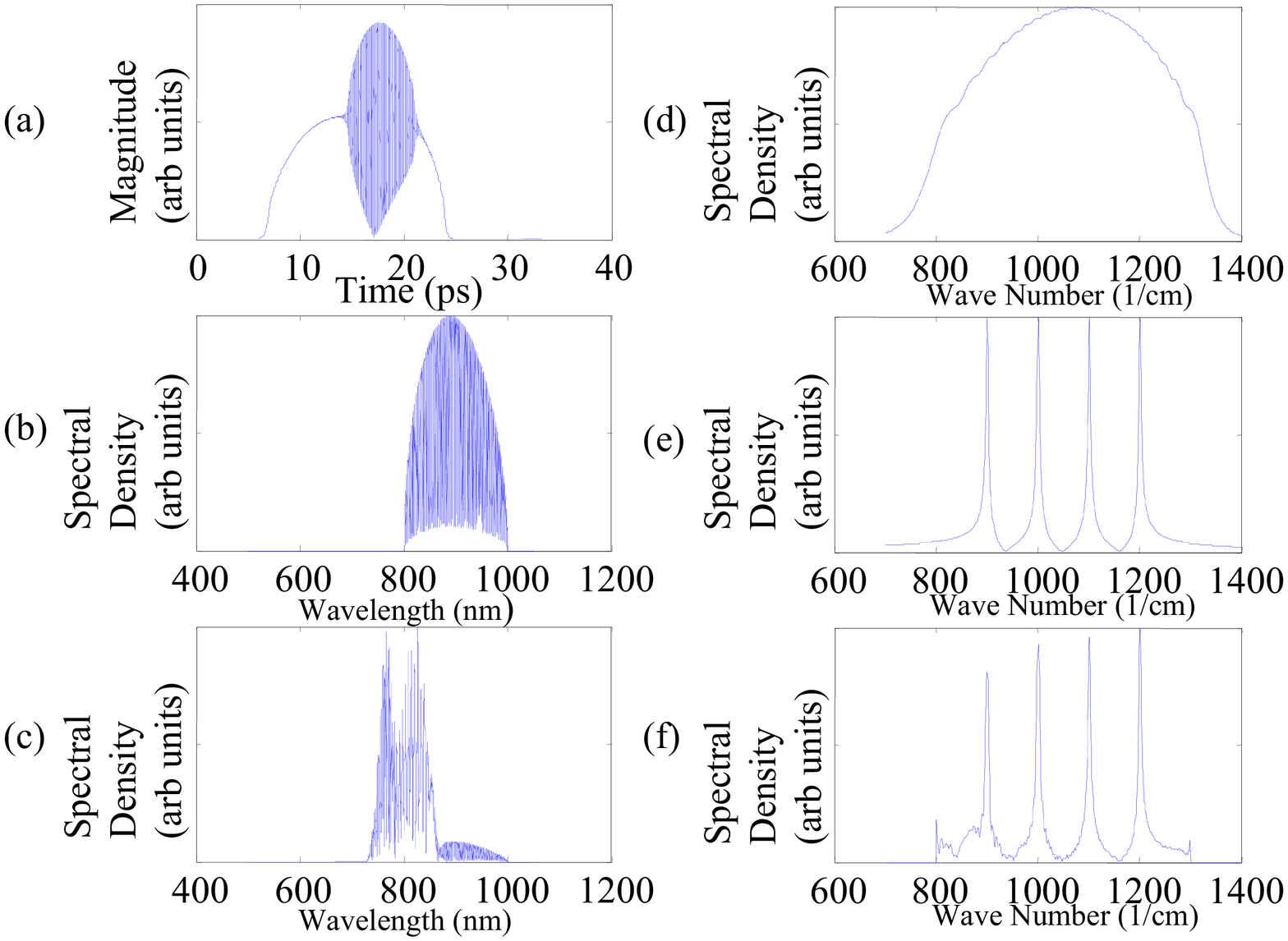,width=90mm,height=79mm} \caption{NIVI
input and output pulses, and the original and reconstructed Raman
spectra in first simulation.  (a) Temporal shape of amplitude of
input pulse.  (b) Power spectral density of input pulse.  (c)
Power spectral density of output pulse.   (d) Power spectrum of
beat frequencies of input pulse.  (e) Magnitude of Raman
susceptibility of hypothetical molecule.  (f) Magnitude of
least-squares reconstructed Raman susceptibility of hypothetical
molecule.} \label{fig:fig3}
\end{figure}

The left column of Figure~\ref{fig:fig3} shows the temporal and spectral shapes of the
input pulse.  Part (a) shows two chirped pulses that partially
overlap, producing a beat pattern that stimulates the resonance.
Part (b) shows the power spectrum of the input pulse.  Part (c)
shows the anti-Stokes radiation spectrum, that is calculated
using Eqs.~\ref{eq:cars1} and~\ref{eq:cars2a}.  Because the
excitation light is assumed to be much more powerful that the
anti-Stokes light, we filter out all of the excitation bandwidth
and utilize only wavelengths shorter than 800 nm for the inverse.
The right column of Figure~\ref{fig:fig3} shows the original and reconstructed $
\chi^{(3)}(\Omega) $.  Part (d) is the magnitude of the spectrum of the
intensity of the original pulse, \emph{i.e.} the beat frequencies of
the pulse.  It shows the possible measurable Raman frequencies with
this pulse.  Part (e) shows the original $ \chi^{(3)}(\Omega) $ Raman
spectrum magnitude.  Finally, part (f) is the Tihkonov- regularized
least-squares reconstructed $ \chi^{(3)}(\Omega) $ based on only the
anti-Stokes frequencies from 700--800 nm.  In simulation, all of the
spectral lines can be recovered.  The minimum discernible separation
in Raman frequencies tends to increase as the Raman frequency
decreases because the anti-Stokes radiation created by lower frequency
resonances tends to overlap the original spectrum more.

\begin{figure}
\epsfig{figure=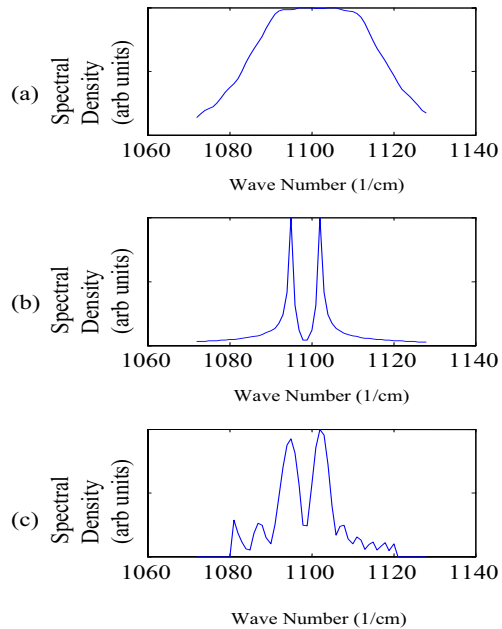,width=90mm,height=99mm}
\caption{Original and reconstructed Raman spectra in second
simulation.  (a) Power spectrum of beat frequencies of input
pulse.  (b) Magnitude of Raman susceptibility of hybrid DNA/RNA
sample.  (c) Magnitude of least-squares reconstructed Raman
susceptibility of hybrid DNA/RNA sample.} \label{fig:fig6}
\end{figure}

As a second demonstration with two closely spaced Raman lines, we
consider deoxyribonucleic acid (DNA), which would be contained in
the nucleus of a cell, and ribonucleic acid (RNA) located
throughout the cell.  Both macromolecules have $ \mbox{PO}_{2} $
phosphodiester resonances, but the resonance occurs in DNA at $
1094\mbox{ cm}^{-1} $ and in RNA at $ 1101\mbox{ cm}^{-1} $.  To
show that a properly designed pulse can recover both resonances
distinctly, we design a $ \chi^{(3)}(\Omega) $ that has resonances
both at $ 1094\mbox{ cm}^{-1} $ and $ 1101\mbox{ cm}^{-1} $, that
could be created by mixing DNA and RNA.  To probe this mixture, we
create a pulse using Eq.~\ref{eq:nivi2} with $ \Omega_L =
1070\mbox{ cm}^{-1} $, $ \Omega_H = 1130\mbox{ cm}^{-1} $, and $ T
= 5\mbox{ ps} $.  The results of this simulation are shown in
Figure~\ref{fig:fig6}. Figure~\ref{fig:fig6} shows the beat
frequency spectrum, and original and reconstructured $
\chi^{(3)}(\Omega) $ Raman spectra.  While the reconstructed lines
are broadened, they are still quite distinct and would be useful
for discerning the two molecules.

NIVI is a flexible tool utilizing ultrafast pulses that can measure
small or large portions of a Raman spectrum of a molecule in a single
pulse.  It does so by interferometrically measuring the anti-Stokes
radiation from a molecule, stimulated by beats in intensity of an
excitation field.  From this anti-Stokes field, the complex Raman
susceptibility can be estimated.  It is especially suited to
biological imaging because while the pulse energy can be large, the
peak power can remain small by chirping the pulse.  For these
reasons, we believe that NIVI can be a general tool for noninvasively
probing the molecular content of biological tissues.

\section{Acknowlegements}
\label{sec:acknowledgement}
We acknowledge the scientific contributions and advice from
Jeremy Bredfeldt, Selezion Hambir, Claudio Vinegoni, Martin
Gruebele, Dana Dlott, Amy Wiedemann, and Barbara Kitchell from
the University of Illinois at Urbana-Champaign.  This research
was supported in part by the National Aeronautics and Space
Administration (NAS2-02057), the National Institutes of Health
(National Cancer Institute), and the Beckman Institute for
Advanced Science and Technlogy.

\end{document}